\documentclass[useAMS,usenatbib]{mn2e}
\usepackage{amsmath}
\usepackage{epsfig}
\usepackage{txfonts}

\usepackage[pdftex,pdfpagemode={UseOutlines},bookmarks,bookmarksopen,colorlinks,linkcolor={blue},citecolor={blue},urlcolor={red}]{hyperref}
\usepackage{float}
\usepackage{graphicx}
\usepackage{caption}
\usepackage{subcaption}
\usepackage{natbib}
\usepackage{rotating,longtable,lscape}
\usepackage{multirow}
\usepackage{pifont}
\title[Circumstellar planets in binaries and exomoons]{Transit probability of precessing circumstellar planets in binaries and exomoons}

\author[Martin]
{\parbox{\textwidth}{David. V. Martin}
\vspace{0.4cm}\\
\parbox{\textwidth}{Observatoire de Gen\`eve, Universit\'e de Gen\`eve, 51 chemin des Maillettes, Sauverny 1290, Switzerland, E-mail: david.martin@unige.ch}}

\begin{document}

\date{Accepted . Received}

\pagerange{\pageref{firstpage}--\pageref{lastpage}} \pubyear{2016}

\maketitle

\label{firstpage}

\begin{abstract}

Over two decades of exoplanetology have yielded thousands of discoveries, yet some types of systems are yet to be observed. Circumstellar planets around one star in a binary have been found, {\it but} not for tight binaries ($\lesssim 5$ AU). Additionally, extra-solar moons are yet to be found. This paper motivates finding both types of three-body system by calculating analytic and numerical  probabilities for all transit configurations, accounting for any mutual inclination and orbital precession. The precession and relative three-body motion  can increase the transit probability  to as high as tens of per cent, and make it inherently time-dependent  over a precession period  as short as 5-10 yr.   Circumstellar planets in such tight binaries present a tempting observational challenge: enhanced transit probabilities but with a quasi-periodic signature that may be difficult to identify. This may help explain their present non-detection, or maybe they simply do not exist. Whilst this paper considers binaries of all orientations, it is demonstrated how eclipsing binaries favourably bias the transit probabilities, sometimes to the point of being guaranteed. Transits of exomoons exhibit a similar behaviour under precession, but unfortunately only have one star to transit rather than two.

\end{abstract}


\section{Introduction}
\label{sec:intro}
We know roughly every star hosts at least one planet \citep{petigura13}, and  that $\sim50\%$ of stars exist in binaries or higher-order multiples \citep{duquennoy91}. It is with natural inquisition that we ponder the existence of planets in multi-star systems, a concept dating back as far as \citet{flammarion74}. For very close binaries ($\lesssim 0.5$ AU) planets are known on wider orbits around both stars - a circumbinary or p-type planet (e.g. Kepler-16, \citealt{doyle11}). In this paper we consider the alternative: a circumstellar or s-type planet that orbits just one of the two stars in a wider binary. Observations have uncovered an influence of the binary separation; stellar companions closer than $\lesssim 100$ AU reduce the occurrence  rate of planets whilst  planets in wider binaries seemingly have a distribution similar to that around single stars \citep{eggenberger07}. The tightest binary known to host a circumstellar planet is the 5.3 AU KOI-1257 \citep{santerne14}\footnote{ Although one may also consider the borderline case of WASP-81. It  contains a hot-Jupiter orbiting a Solar mass star, with an outer 2.4 AU Brown Dwarf companion with a minimum mass of 56.6 $M_{\rm Jup}$ \citep{triaud16}.}, and no circumstellar planets are known in eclipsing binaries.

 The observed paucity of circumstellar planets in tight binaries does have theoretical merit \citep{kraus16}. Each star is expected to have its own circumstellar disc, but if the two stars are close enough this will be tidally truncated to roughly 15-35\% of the binary separation \citep{artymowicz94}. Binaries closer than $\sim 10-30$ AU may even truncate the disc interior to its snow line, suggesting that giant planet formation is inhibited \citep{kraus12}, although one must recall that the previously mentioned KOI-1257 5.3 AU binary hosts a 1.45 $M_{\rm Jup}$ planet.

 However, despite the paucity of circumstellar planets in tighter binaries,  one must remember that the history of exoplanets is one of surprises.  The discovery of even a single circumstellar planet orbiting within a very tight binary would already pose intriguing theoretical questions, whilst a sample of many would revolutionise our understanding of the robustness of planet formation. One should therefore be {\it motivated}, rather than deterred, to find such planets. The recent paper \citet{oshagh16} followed this philosophy,  assessing the detectability of circumstellar planets orbiting close eclipsing binaries using a combination of radial velocities and  eclipse timing variations. This alleviates degeneracies between the planet mass and semi-major axis that exist when using solely eclipse timing variations. 

The work of \citet{oshagh16} has roots in a conceptually similar three-body system: extra-solar moons. Methods to detect them have focused on the transit timing and duration variations they induce on a transiting exoplanet \citep{sartoretti99,kipping09} and transits of the moons themselves \citep{sartoretti99}. However, despite search attempts using the high  precision and lengthy continuous photometry of {\it Kepler} \citep{kipping15}, detections have not yet been forthcoming.  In the Solar System no moons are known around planets less than 1 AU from the Sun, but another lesson from two decades of exoplanet surveys has been to avoid neighbourhood-induced preconceptions.

This paper is an analytic and numerical study of the transit probability of both circumstellar planets in binaries and exomoons   in planetary systems. The mathematics and geometry derived are done in full generality to be equally applicable to both cases. Advances are made on existing studies by including Newtonian three-body orbital precession of the planet/moon and considering all mutual inclinations.  This is shown to both increase the transit probability and make it time-dependent, in a way similar to circumbinary planets \citep{schneider94,martin14,martin15,martin17}. The work is applicable to all binaries, not just those which eclipse.

The  geometry and orbital dynamics are introduced in Sect.~\ref{sec:setup}. Next, in Sect.~\ref{sec:host_body} we derive the probability of transiting the host body that the satellite (planet or moon) orbits, before in Sect.~\ref{sec:comp_body} deriving the probability of transiting the other, companion body. Finally in Sect.~\ref{sec:applications} N-body simulations are run to test a variety of examples and illustrate the observational signature, before concluding.

\section{Problem Setup}\label{sec:setup}


\begin{figure}  
\begin{center}  
\includegraphics[width=0.49\textwidth]{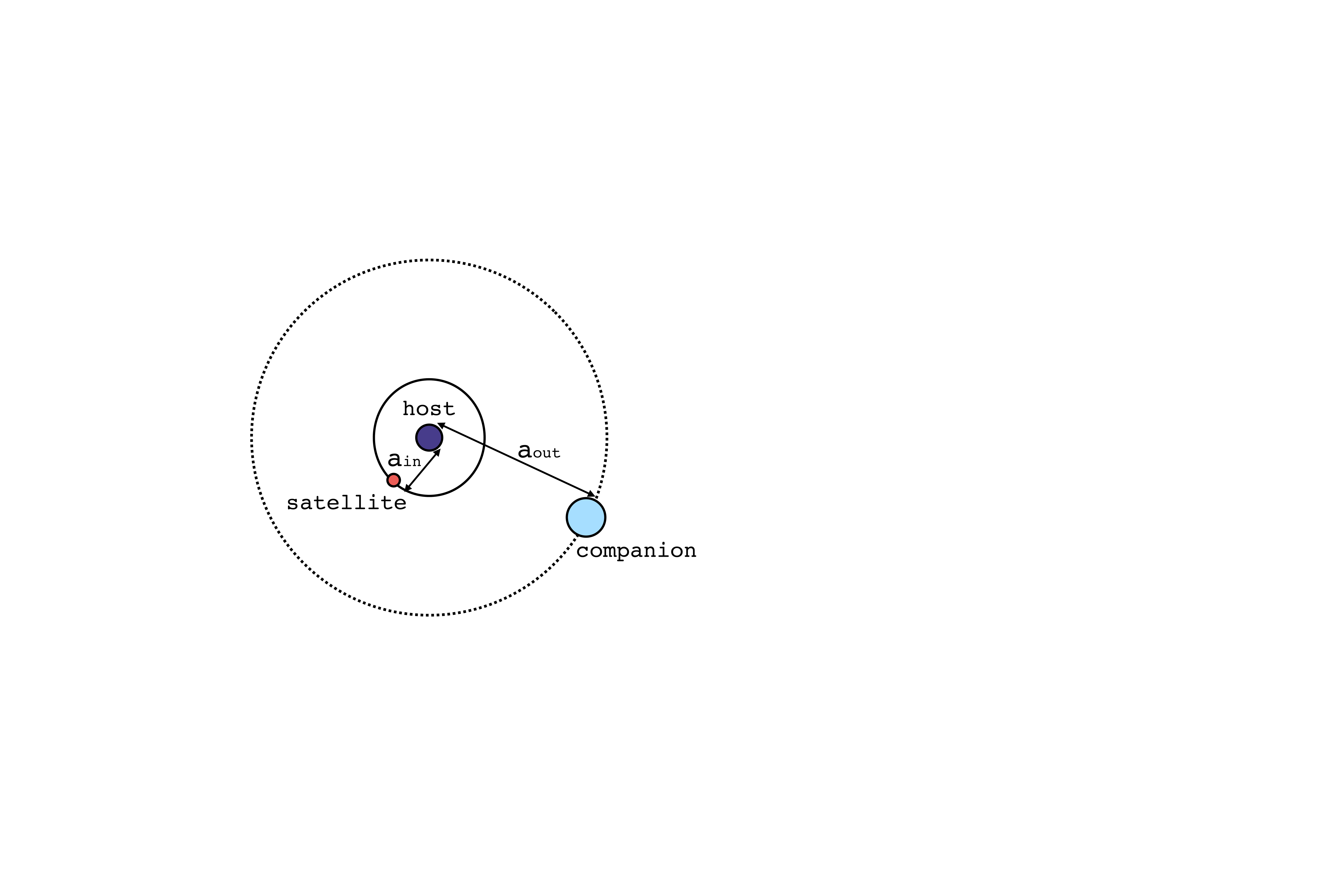}  
\caption{Geometry of the inner restricted three-body problem. For circumstellar planets in binaries the red massless satellite is the planet and the two blue bodies are stars. In this image the planet is arbitrarily orbiting the secondary star. For exomoons the red massless satellite is the moon, the dark blue host is the planet and the light blue companion is the star.}
\label{fig:base_geometry}
\end{center}  
\end{figure} 

We consider  the inner restricted three-body problem,  illustrated in Fig.~\ref{fig:base_geometry}. Namely, there is a massless satellite  (the circumstellar planet in a binary or the exomoon) on a close orbit of period $T_{\rm in}$ and semi-major axis $a_{\rm in}$ around a host body  (for the exomoon the host is the planet) of mass $M_{\rm host}$ and radius $R_{\rm host}$. On a wider orbit ($T_{\rm out}$, $a_{\rm out}$) is an outer companion  (for the exomoon the companion is the star) with $M_{\rm comp}$ and $R_{\rm comp}$. It is an ``inner restricted" problem because the outer orbit contains 100\% of the angular momentum.  The transit probabilities derived in this paper would also work for a more massive satellite as long as the inner orbit's angular momentum remained much less than the outer. The orientation of each orbit with respect to the observer is characterised by the inclination $I$ and the longitude of the ascending node $\Omega$. The orbital dynamics between the two orbits are dictated by the relative orientation, which we define using the mutual inclination,

\begin{equation}
\label{eq:Delta_I}
\cos \Delta I = \cos \Delta \Omega\sin I_{\rm in}\sin I_{\rm out} + \cos I_{\rm in} \cos I_{\rm out},
\end{equation}
where $\Delta \Omega = \Omega_{\rm in} - \Omega_{\rm out}$. Transit observations are only sensitive to $\Delta \Omega$ and not the individual values of $\Omega$ so we arbitrarily set $\Omega_{\rm out}=0$. Both orbits are assumed to be circular.

\begin{figure}  
\begin{center}  
\includegraphics[width=0.49\textwidth]{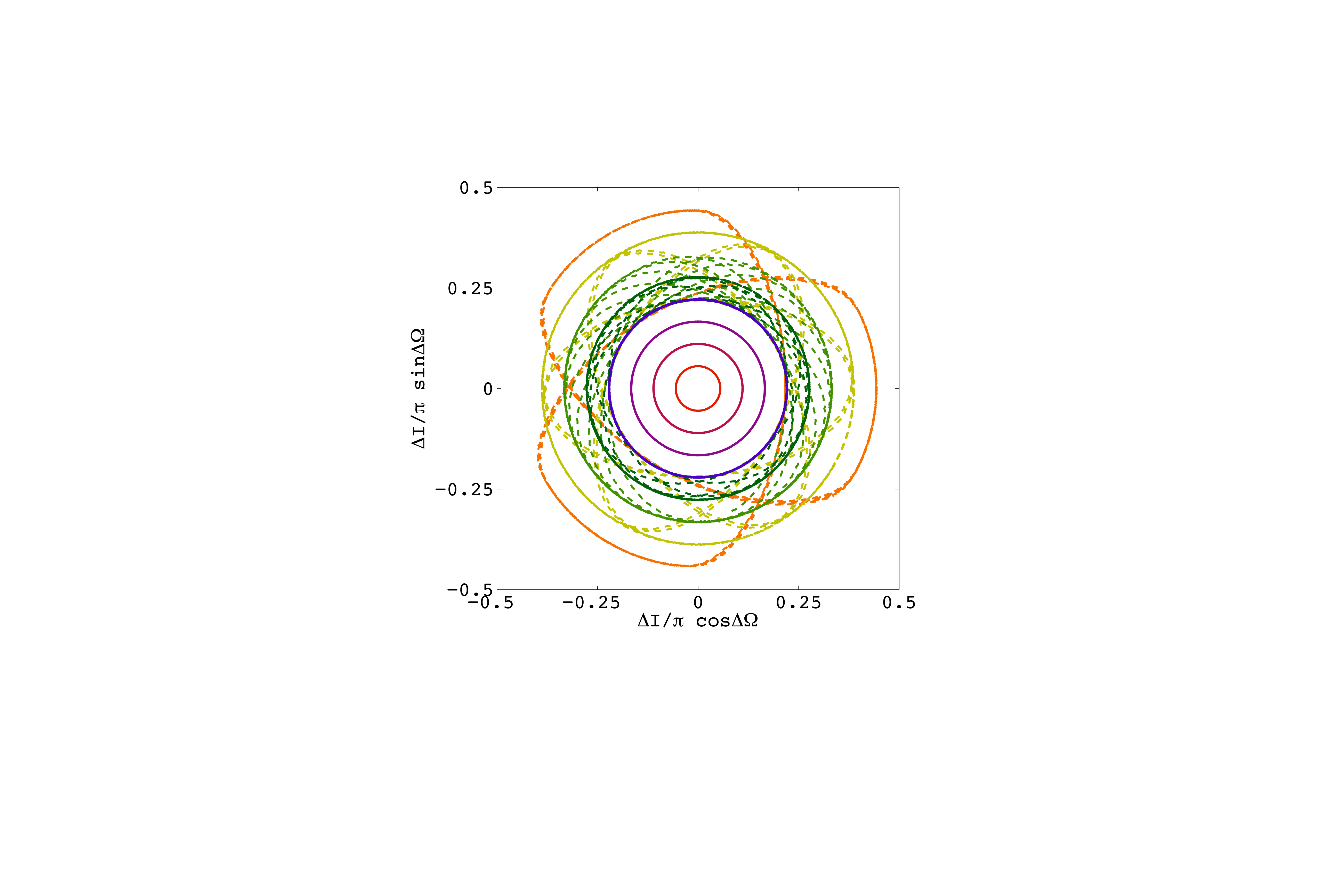}  
\caption{ N-body simulations over 5,000 years of a massless circumstellar planet and two Solar stars with $a_{\rm in}=0.1$ AU, $a_{\rm out} = 2$ AU  and different initial $\Delta I$. Systems devoid of large Kozai-Lidov cycles are the central circles drawn with sold lines, where $\Delta I_0$ ranges from $10^{\circ}$ (red) to $40^{\circ}$ (purple) in steps of $10^{\circ}$. Systems with Kozai-Lidov cycles have a more complicated evolution and are drawn with dashed lines for $\Delta I_0$ ranging from $50^{\circ}$ (dark green) to $80^{\circ}$ (orange). In all simulations $\Delta \Omega_0=0^{\circ}$.}
\label{fig:kozai_precession}
\end{center}  
\end{figure} 

\begin{figure}  
\begin{center}  
	\begin{subfigure}[b]{0.49\textwidth}
		\caption{Fixed $M_{\rm comp}=1M_{\odot}$, variable $M_{\rm host}$}
		\includegraphics[width=\textwidth]{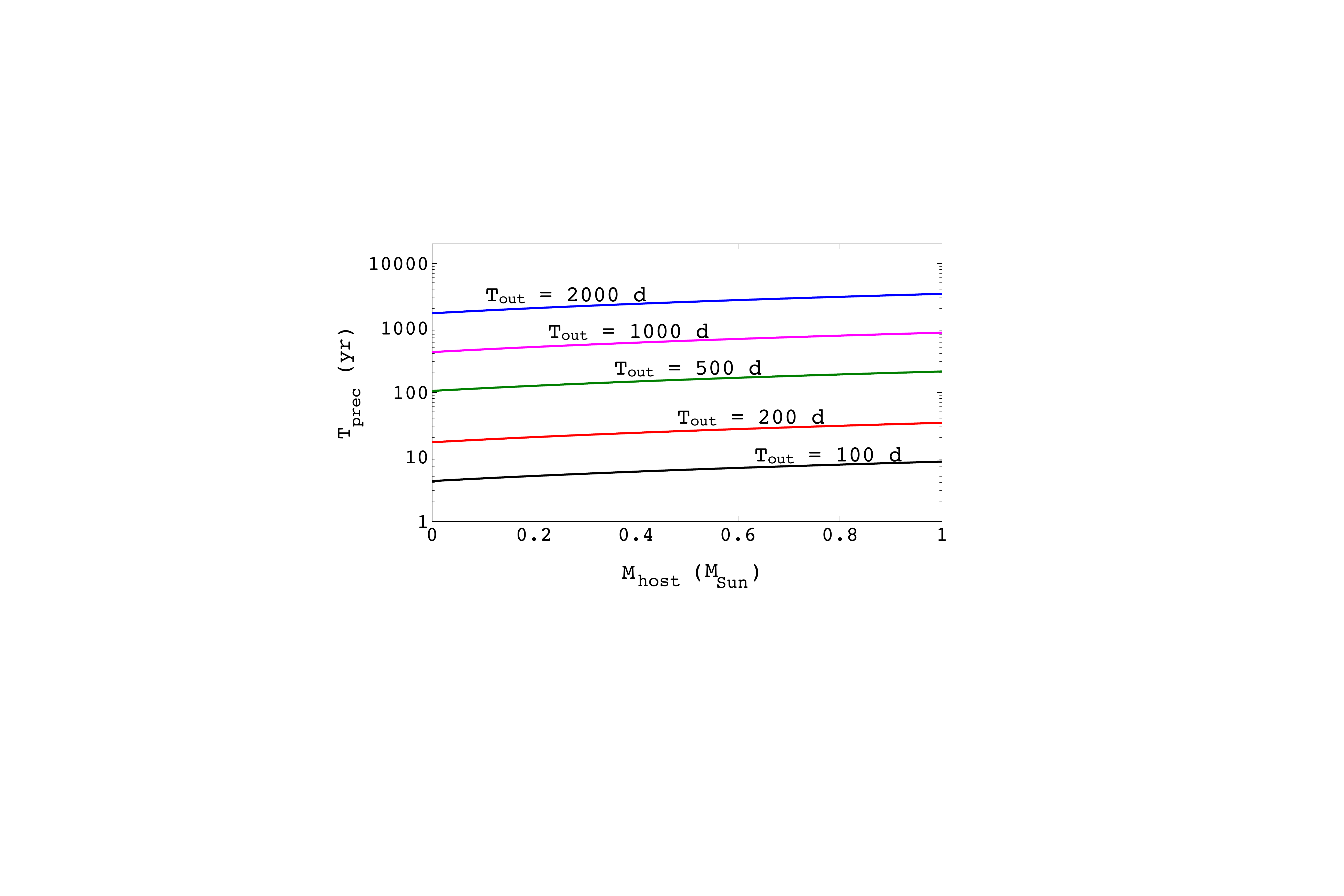}  
	\end{subfigure}
	\medskip
	\begin{subfigure}[b]{0.49\textwidth}
		\caption{Fixed $M_{\rm host}=1M_{\odot}$, variable $M_{\rm comp}$}
		\includegraphics[width=\textwidth]{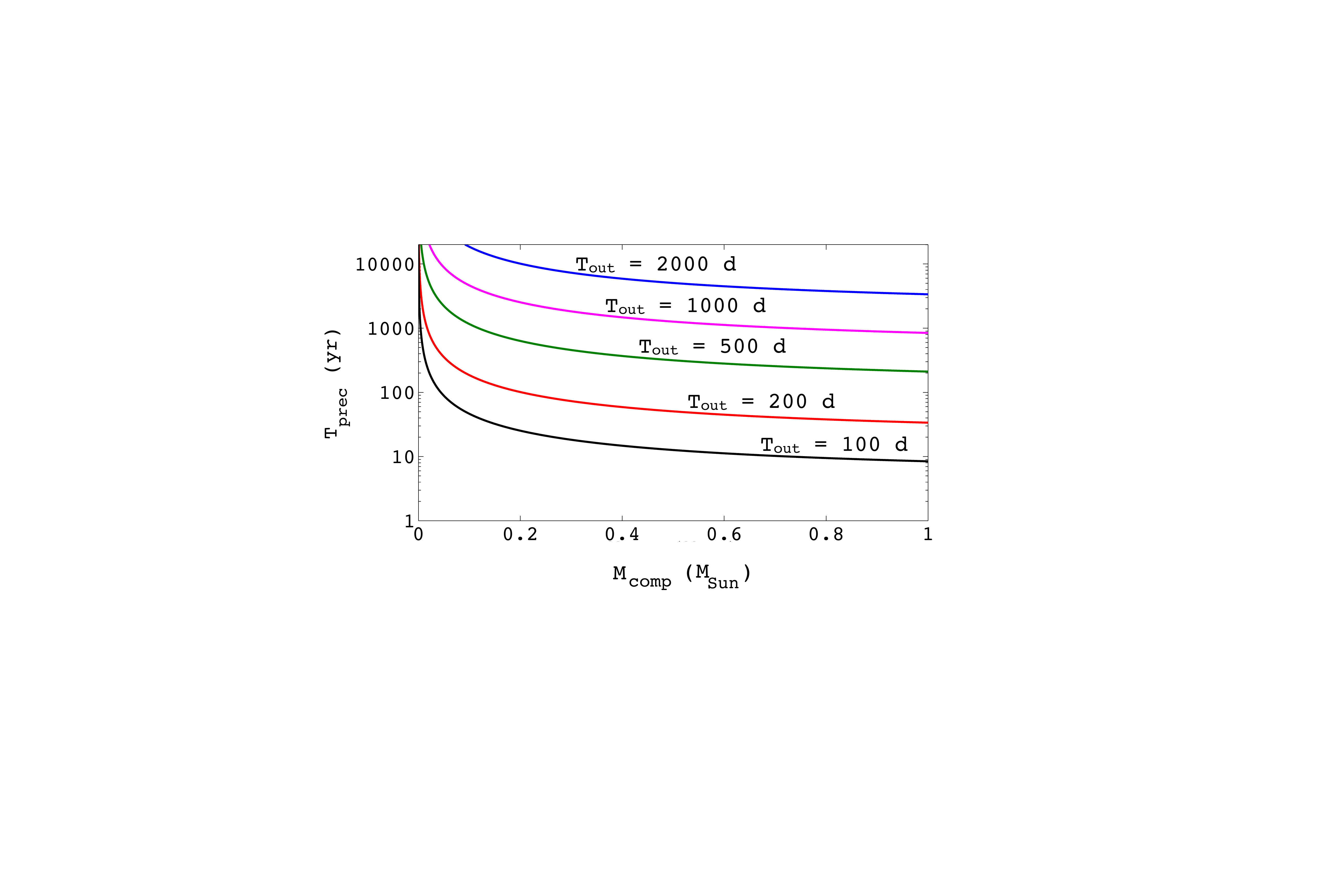}  	
	\end{subfigure}
	\caption{Precession period calculated by Eq.~\ref{eq:precession_period} where in all simulations $T_{\rm in}=10$ d and $\Delta I=30^{\circ}$, whilst $T_{\rm out}=100$ d (black), 200 d (red), 500 d (green), 1000 d (magenta) and 2000 d (blue). In (a) $M_{\rm comp}$ is fixed at $1M_{\odot}$ and $M_{\rm host}$ varies between 0 (like for an exomoon) and $1M_{\odot}$ (like for a circumstellar planet in a binary). In (b) instead $M_{\rm host}$ is fixed at $1M_{\odot}$ and $M_{\rm comp}$ varies between 0 (like for a single star planet with an outer planetary companion) and $1M_{\odot}$ (like for a circumstellar planet in a binary).}
\label{fig:Prec_Period_Example}  
\end{center}  
\end{figure}

The inner orbit feels gravitational perturbations from the outer orbit and consequently evolves with time,  whilst the outer orbit remains constant. In Fig.~\ref{fig:kozai_precession} we illustrate the evolution of the orientation of the inner orbit with respect to the outer, using N-body simulations of a 0.1 AU planet orbiting one star in a 2 AU binary of two Solar stars, with different mutual inclinations. It is seen that for  $\Delta I \lesssim 40^{\circ}$ there is a precession of $\Delta \Omega$ over a full $360^{\circ}$, whilst $\Delta I$ remains constant (the evolution traces out a circle in the $\Delta I \cos \Delta \Omega$ vs $\Delta I \sin \Delta \Omega$ domain). If $\Delta I$ is initially greater than $\gtrsim 40^{\circ}$ then the precession behaviour is different under the influence of the Kozai-Lidov effect \citep{lidov61,kozai62}, in which case $\Delta I$ begins to vary along with $e_{\rm in}$, even for initially circular orbits. Constancy of $\Delta I$ is fundamental in this paper so we are restricted to $\Delta I\lesssim40^{\circ}$.  Assuming this, the observational consequence of this precession is that $I_{\rm in}$ oscillates around the constant $I_{\rm out}$,

\begin{equation}
\label{eq:I_in}
I_{\rm in}(t) = \Delta I \cos\left(\frac{2\pi t}{T_{\rm prec}}\right) + I_{\rm out},
\end{equation}
where $t$ is time, $T_{\rm prec}$ is the precession period and the semi-amplitude of oscillation $\Delta I$ is constant. Equation~\ref{eq:I_in} assumes  $\Omega_{\rm in}=0$ at $t=0$.

 The rate of precession was calculated by \citet{mardling10} to be
\begin{align}
\label{eq:precession_period}
T_{\rm prec}= \frac{4}{3}\frac{M_{\rm host}+M_{\rm comp}}{M_{\rm comp}}\frac{T_{\rm out}^2}{T_{\rm in}}\frac{1}{\cos \Delta I},
\end{align}
where both orbits are assumed to be circular. In Fig.~\ref{fig:Prec_Period_Example} we evaluate Eq.~\ref{eq:precession_period} for $T_{\rm in}=10$ d, $T_{\rm out}$ between 100 and 2000 d, $\Delta I = 30^{\circ}$ and various $M_{\rm host}$ and $M_{\rm comp}$ between 0 and $1M_{\odot}$. In Fig.~\ref{fig:Prec_Period_Example}a we see that there is only a weak dependence on $M_{\rm host}$, and that precession of an exomoon is slightly faster than of a circumstellar planet in a binary. This contrasts with Fig.~\ref{fig:Prec_Period_Example}b where $T_{\rm prec}$ becomes very large as $M_{\rm comp}$ as it goes to zero. In the examples shown $T_{\rm prec}$ may be as short as 5-10yr, which is roughly comparable to the original {\it Kepler} mission's lifetime. The nodal precession of the Moon around the Earth takes 18 yr, where $\Delta I = 5.14^{\circ}$.


%

\begin{figure}  
\begin{center}  
	\begin{subfigure}[b]{0.49\textwidth}
		\caption{}
		\includegraphics[width=\textwidth]{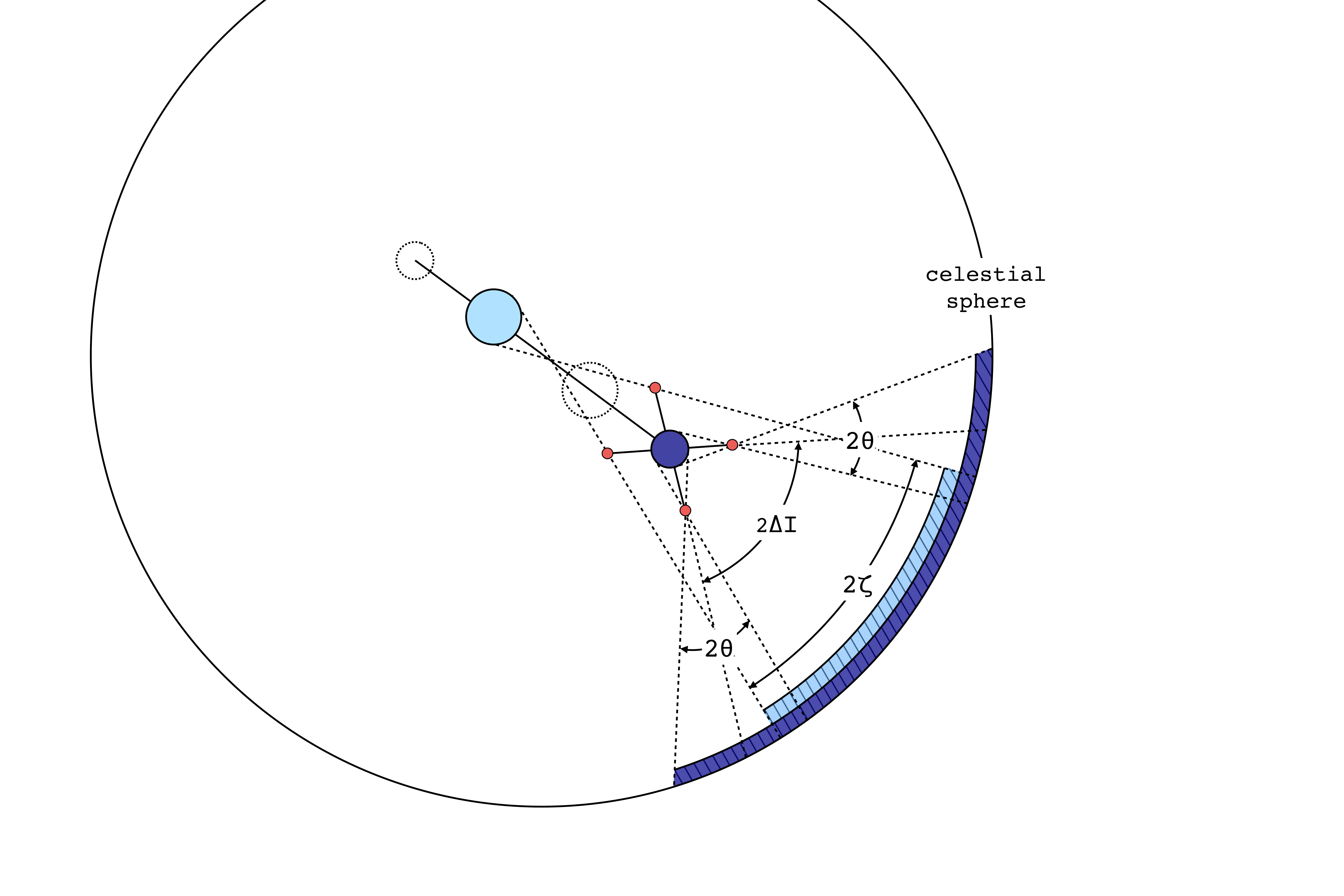}  
	\end{subfigure}
	\begin{subfigure}[b]{0.49\textwidth}
		\caption{}
		\includegraphics[width=\textwidth]{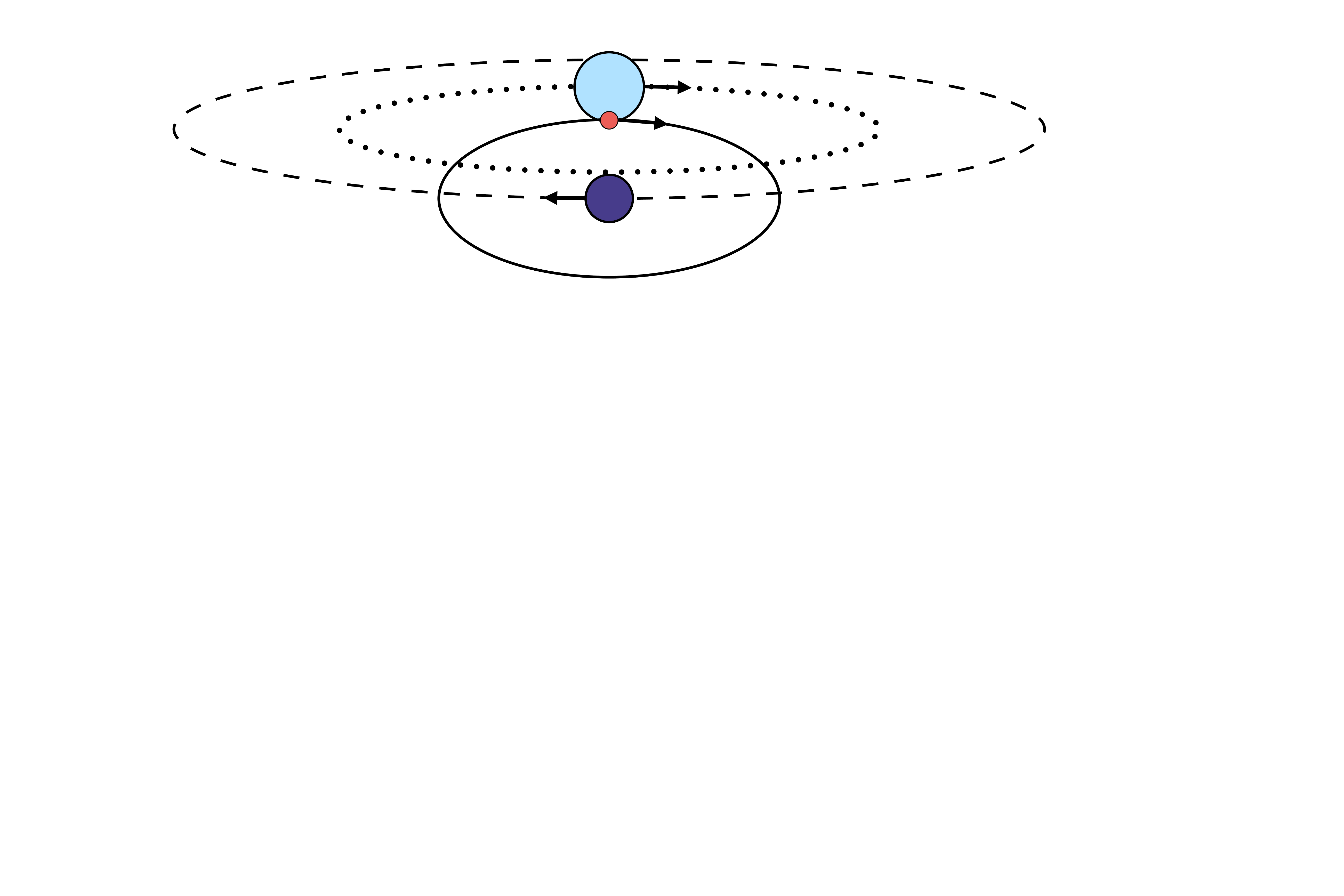}  	
	\end{subfigure}
	\begin{subfigure}[b]{0.49\textwidth}
		\caption{}
		\includegraphics[width=\textwidth]{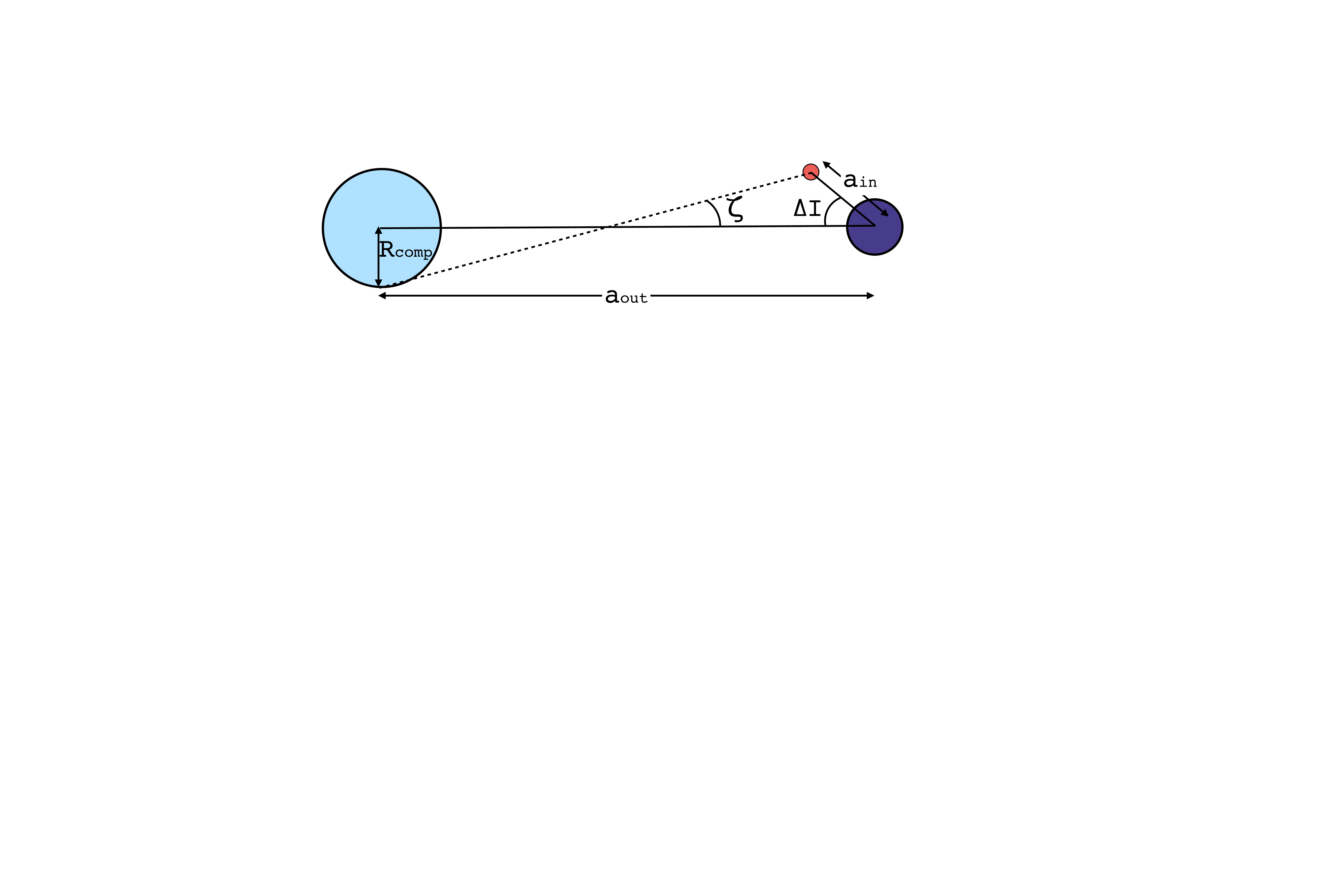}  
	\end{subfigure}
	\caption{(a) Side-on view of a satellite (red) orbiting its host (dark blue) which is in turn in orbit around the companion body (light blue). The satellite orbit varies during precession, and is illustrated in the two extreme orientations. Dark and light blue hatched regions projected onto the celestial sphere correspond to observers who will see transits on the host and companion body, respectively. (b) Front-on view of an example transit on the companion. (c) Zoomed side-on view used in deriving Eq.~\ref{eq:prob_comp_body}.}
\label{fig:geometry}  
\end{center}  
\end{figure} 

\section{Probability to transit the host}\label{sec:host_body}

For a circumprimary/circumsecondary planet the host body being orbited is a star. For an exomoon, however, the host body is the planet and the photometric signal of a moon-planet transit is likely beyond present detection capabilities, but \citet{cabrera07} remain optimistic. Regardless, the geometry, mathematics and notation used are applicable to both types of system.

The chance of the satellite on the inner orbit transiting its host is solely dictated by their relative orientation; the movement of the host in the outer orbit is irrelevant. The transit criterion is 

\begin{equation}
\left| I_{\rm in}-\frac{\pi}{2}\right| < \theta = \sin^{-1}\left(\frac{R_{\rm host}}{a_{\rm in}}\right). 
\end{equation}
If this orientation were constant then the probability of transiting the host body would be simply akin to that of a planet around a single star: $P_{\rm host}=R_{\rm host}/a_{\rm in}$. However,  in close binaries one must account for variation of the satellite's orbital orientation. In Fig.~\ref{fig:geometry}a we illustrate a host body (dark blue) and companion body (light blue) moving back and forth along the solid line connecting them, with dotted circles indicating their positions at the opposite edge of their orbit. The red satellite orbits its host body with two orientations drawn, corresponding to the extremes of its inclination, $I_{\rm in} = I_{\rm out} \pm \Delta I$, which it reaches during its  precession period. The satellite orbit subtends an angle of $2\Delta I$ on the celestial sphere. This $\Delta I$, in addition to the size of the host body relative to the satellite, $\theta = \sin^{-1}(R_{\rm host}/a_{\rm in})$, defines the region of observers on the celestial sphere (dark blue hatched region) that will see a transit on the host at some point during the precession period.

To calculate the probability that the satellite will transit its host at some unspecified point in time is to calculate the probability of an observer being within the dark blue hatched region in Fig.~\ref{fig:geometry}a.  We assume that the three body system is randomly oriented on the celestial sphere, which is characterised by a uniform distribution of  $\cos I_{\rm out}$, and hence a probability density function of $\sin I_{\rm out}$. We then integrate $\sin I_{\rm out}$ over the angle $\Delta I + \theta$:
\begin{align}
\begin{split}
\label{eq:prob_host_body}
P_{\rm host} &= \int_{\pi/2 - \Delta I - \theta }^{\pi/2}\sin I_{\rm out} dI_{\rm out}, \\
&= \sin\left[\Delta I + \sin^{-1}\left(\frac{R_{\rm host}}{a_{\rm in}}\right)\right].
\end{split}
\end{align}
 The integral is just done over a quarter of the projected celestial sphere ($\pi/2$) but symmetry permits this. A large $\Delta I$ increases $P_{\rm host}$ as the satellite subtends a greater range of angles on the sky. Like in the single star case, orbiting close to a large host body increases the likelihood. For $\Delta I$ beyond a few degrees transits will not be continual but rather will come and go over time as  $I_{\rm in}$ changes. However, {\it if} the satellite is ever to transit then it is guaranteed to do so within a single precession period. 

%


\section{Probability to transit the companion}\label{sec:comp_body}

If both orbits were perfectly aligned with the observer ($I_{\rm out}=I_{\rm in}=90^{\circ}$) then the satellite would trivially transit both the host and companion. However, transits on the companion are still possible for inclined cases. In Fig.~\ref{fig:geometry}b is an example, showing a front-on observer perspective of a red satellite orbiting the dark blue host, transiting the light blue companion. However, unlike in Sect.~\ref{sec:host_body} where the host body is stationary with respect to the satellite, the companion  is moving. This means that overlapping orbits of the satellite and companion, accounting for its radius, make transits possible but not guaranteed on every passing. We call such a configuration ``transitability" as was done in \citet{martin14} to describe the conceptually similar case of inclined circumbinary planets. \citet{sartoretti99} call this the ``geometric probability."

To work out if transits are possible we calculate the limits of transitability, corresponding to when the projected separation of the stars is minimised like in Fig.~\ref{fig:geometry}b. The satellite, whose orbit is tied to its host, must ``reach" over to the companion  to transit. This reach is maximised when $\Omega_{\rm in}=0$\footnote{$\Omega_{\rm in}$ rotates the projected satellite orbit counter-clockwise.}. During the precession period $\Omega_{\rm in}=0$ (and hence $\Delta \Omega=0$ too) at the extremities  $I_{\rm in}=I_{\rm out}\pm \Delta I$.

In Fig.~\ref{fig:geometry}a the region of transitability on the companion body is denoted as a light blue hatched region. Generally this region is smaller than that for the host body. The angle $\zeta$ which characterises this region is defined more clearly in Fig.~\ref{fig:geometry}c and is

\begin{equation}
\zeta = \tan^{-1}\left(\frac{a_{\rm in}\sin \Delta I + R_{\rm comp}}{a_{\rm out}-a_{\rm in}\cos \Delta I}\right).
\end{equation}
Like in Sect.~\ref{sec:host_body} one may integrate over this angle to calculate the probability of transitability:

\begin{align}
\begin{split}
\label{eq:prob_comp_body}
P_{\rm comp} &= \int_{\pi/2 - \zeta }^{\pi/2}\sin I_{\rm out} dI_{\rm out}, \\
&= \sin\left[\tan^{-1}\left(\frac{a_{\rm in}\sin \Delta I + R_{\rm comp}}{a_{\rm out} - a_{\rm in}\cos \Delta I}\right)\right], \\
&\approx \frac{a_{\rm in}\sin \Delta I + R_{\rm comp}}{a_{\rm out} - a_{\rm in}\cos \Delta I}.
\end{split}
\end{align}

As was seen for transits on the host star, a mutual inclination increases the transit probability. There is also the expected result that companion bodies that are larger and closer are more likely to be transited, although the companion cannot be too close otherwise the system will be unstable (e.g. \citealt{holman99}). The efficiency of transitability, i.e. the chance that overlapping orbits lead to transits within a given time, is a function of the various orbital parameters and is not trivial to calculate, as was seen in the case of circumbinary planets \citep{martin15,martin17}. In \cite{sartoretti99} this is called the ``orbital probability" and equations are derived in the case of exomoons but without the inclusion of orbital precession. In this present paper we use numerical methods for this.
\section{Analysis and Applications}\label{sec:applications}

\subsection{Comparison with numerical simulations}\label{sec:examples}

A simple test is run to show the validity of the analytic work. The percentage of systems transiting over 35 years is calculated numerically as a Monte Carlo experiment of 2000 N-body simulations\footnote{A 4th-order Runge-Kutta with a 30 minute time step, which is the standard {\it Kepler} observing cadence and conserves energy to within $\sim10^{-12} \%$. All Monte Carlo simulations run in this paper are done with 2000 randomised systems.} where the three-body orientation with respect to the observer is randomised isotropically.  A system is recorded to have a transit if the projected sky position of the satellite intersects the host or companion disc at least once. The detectability of the transit signal, i.e. its depth, duration and frequency, is not accounted for in the transit probability. However, taking the satellite as a point body is equivalent to demanding at least half of the satellite's disc intersects the host/companion disc, and hence highly grazing transits are excluded. The expected timing of transits  is discussed briefly in Sect.~\ref{subsec:observational_signature}. 

\begin{figure}  
\begin{center}  
\includegraphics[width=0.49\textwidth]{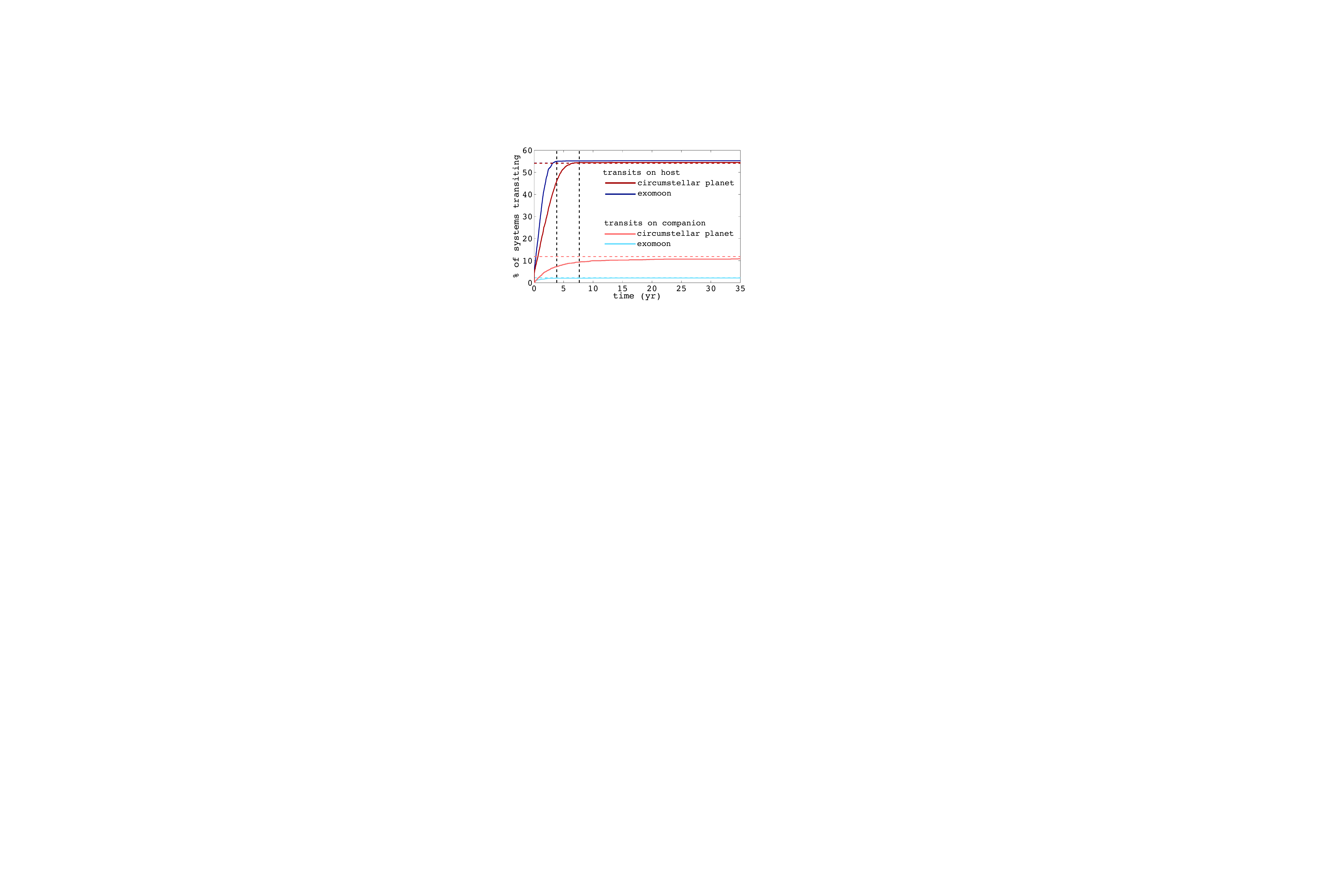}  
\caption{Example N-body simulated transit percentages compared with analytic theory, for a circumstellar planet in a binary ($T_{\rm in}=11$ d, $T_{\rm out}=100$ d, $\Delta I = 30^{\circ}$, two Solar stars) and an exomoon (same parameters but the host is a Jupiter). Horizontal dashed lines are the analytic equations (Eq.~\ref{eq:prob_host_body}, upper and Eq.~\ref{eq:prob_comp_body}, lower) and solid lines are N-body simulations. For the binary case dark and light red are transits on the host and companion bodies, respectively, and for exomoons dark and light blue are host and companion transits. The two black vertical dashed lines indicate the  precession period for the binary case (7.67 yr) and exomoon (3.83 yr). The analytic equation for transits on the host coincidentally gives almost an identical value in both cases, and hence the dark blue and dark red horizontal dashed lines overlap.}
\label{fig:example}
\end{center}  
\end{figure}

The numerical results are then compared with Eqs.~\ref{eq:prob_host_body} and ~\ref{eq:prob_comp_body}. Two example systems are tested: for both $T_{\rm in}=11$ d, $T_{\rm out} = 100$ d, $\Delta I=30^{\circ}$ and the companion body is a Sun. In one system the host body is also a Sun, and in the other it is a Jupiter. Owing to Kepler's laws, $a_{\rm in}=0.0968$ AU for the circumstellar planet and $a_{\rm in}=0.0095$ AU for the exomoon\footnote{An exomoon at 0.0968 AU with all other parameters kept the same would be unstable because the hill sphere of a Jupiter-mass planet is much smaller than a Solar-mass star.}. The angle between the two orbits, $\Delta \Omega$, is randomised between 0 and 360$^{\circ}$, meaning that each phase of the precession period is uniformly sampled. The results are shown in Fig.~\ref{fig:example}. 

In all four cases (i.e. the binary and exomoon cases and for host and companion transits) the analytic theory accurately matches the amount of transits after observing for many years. As predicted, the percentage of systems transiting the host body plateaus in less than a precession period.  The percentage of systems transiting the host star almost instantaneously (within one $T_{\rm in}=11$ d) rises to $R_{\rm host}/a_{\rm in}$, since this is the probability for a static orbit sans precession. Coincidentally this ratio is almost the same in the two systems tested, despite $R_{\rm host}$ and $a_{\rm in}$ being significantly different in the two cases, and hence the dark red and dark blue curves both start at nearly the same percentage at $t=0$ and also plateau at nearly the same percentage according to Eq.~\ref{eq:prob_host_body}. What is different between the two cases is the precession period, with the exomoon precessing faster. The percentage of systems with transits on the companion body  rises slower and continues to do so  after the precession period, as expected, and is significantly higher for the binary case (light red) than the exomoon case (light blue).  This latter result is because $a_{\rm in}$ is much larger in the binary case (0.0968 AU) compared to the exomoon (0.0095 AU).

\subsection{Circumstellar planets in binaries}\label{sec:examples}

\begin{figure}  
\begin{center}  
\large{Circumprimary planet simulations}
\includegraphics[width=0.49\textwidth]{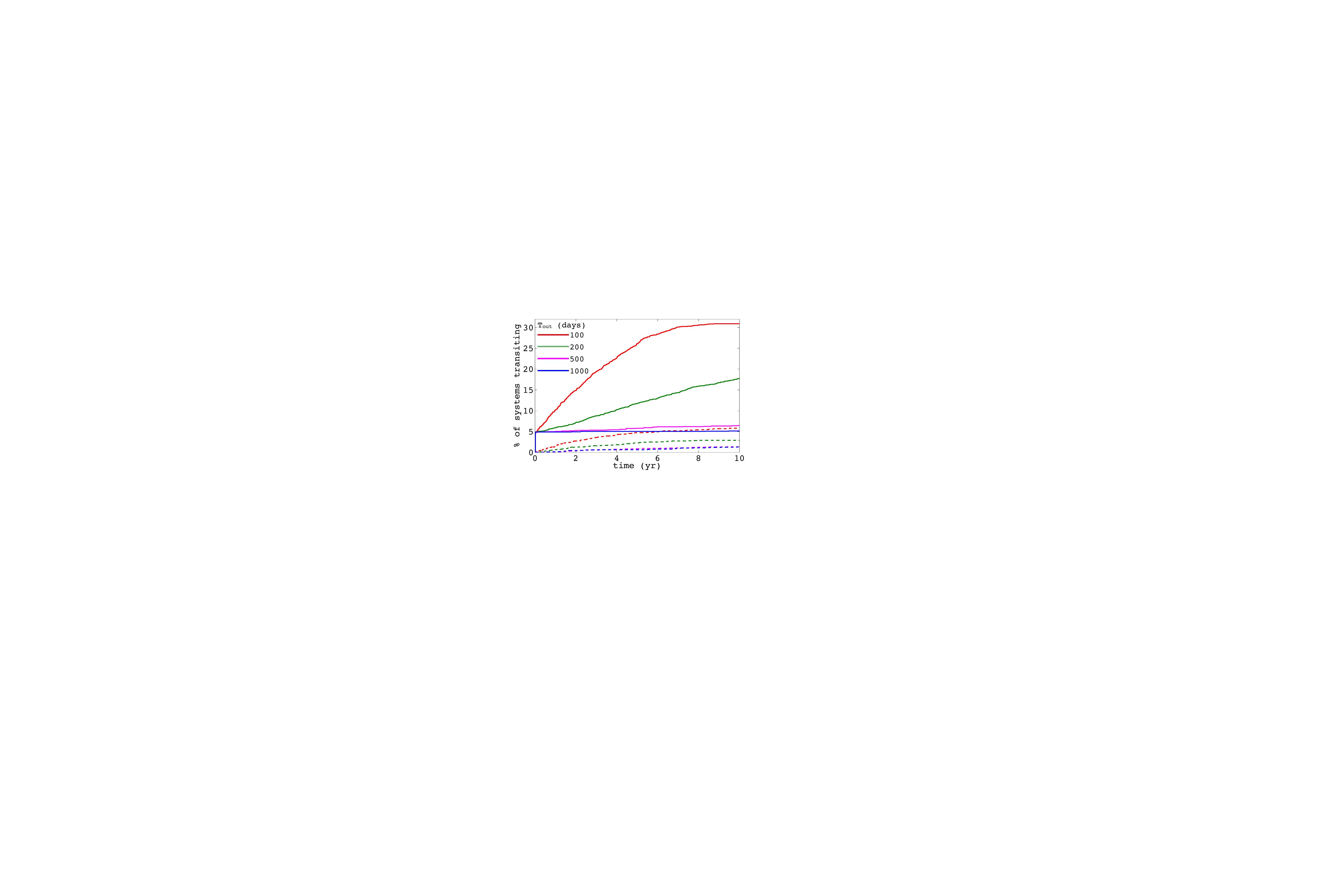}  
\caption{N-body simulations of transits on the host star (solid lines) and companion star (dashed lines) for a circumprimary planet in a binary with Solar and half-Solar stars. In all simulations $T_{\rm in}=11$ d (planet period) and $\Delta I=15^{\circ}$. Different colours indicate different binary periods $T_{\rm out}$: 100 d (red), 200 d (green), 500 d  (magenta) and 1000 d (blue).}
\label{fig:outer_period_test}
\end{center}  
\end{figure}

\subsubsection{Effect of the outer period}

Several tests are run to calculate the time-dependent transit probability of circumstellar planets in binaries. Numerical simulations are run over 10 years to cover the existing {\it Kepler} mission (four years) and what may be considered reasonable for future missions. First, in Fig.~\ref{fig:outer_period_test} we test the effect of the binary period, $T_{\rm out}$, between 100 and 1000 d. The binary is constructed with a Solar primary and half-Solar secondary (in terms of both mass and radius). The planet is arbitrarily chosen to orbit the primary star and has a fixed $T_{\rm in}=11$ d and mutual inclination $\Delta I = 15^{\circ}$. 

Transits on the host star that the planet orbits  initially have a probability of $R_{\rm host}/a_{\rm in}$, which then increases over time due to precession. For $T_{\rm out}=100$ d a complete precession period is completed within the 10 year observing window. At the other extreme, for $T_{\rm out}=1000$ d the precession period is so long that the transit probability barely increases over 10 years. Ultimately, according to Eq.~\ref{eq:prob_host_body} the transit probability will reach the same value regardless of $T_{\rm out}$, but practically, accounting for orbital precession is only important for binaries of a couple of hundred days period. Transits on the companion star are systematically less likely than transits on the host star, and similarly tend to be favoured in tighter binaries.

\subsubsection{Effect of the inner period}

\begin{figure}  
\begin{center}  
\large{Circumprimary planet simulations}
\includegraphics[width=0.49\textwidth]{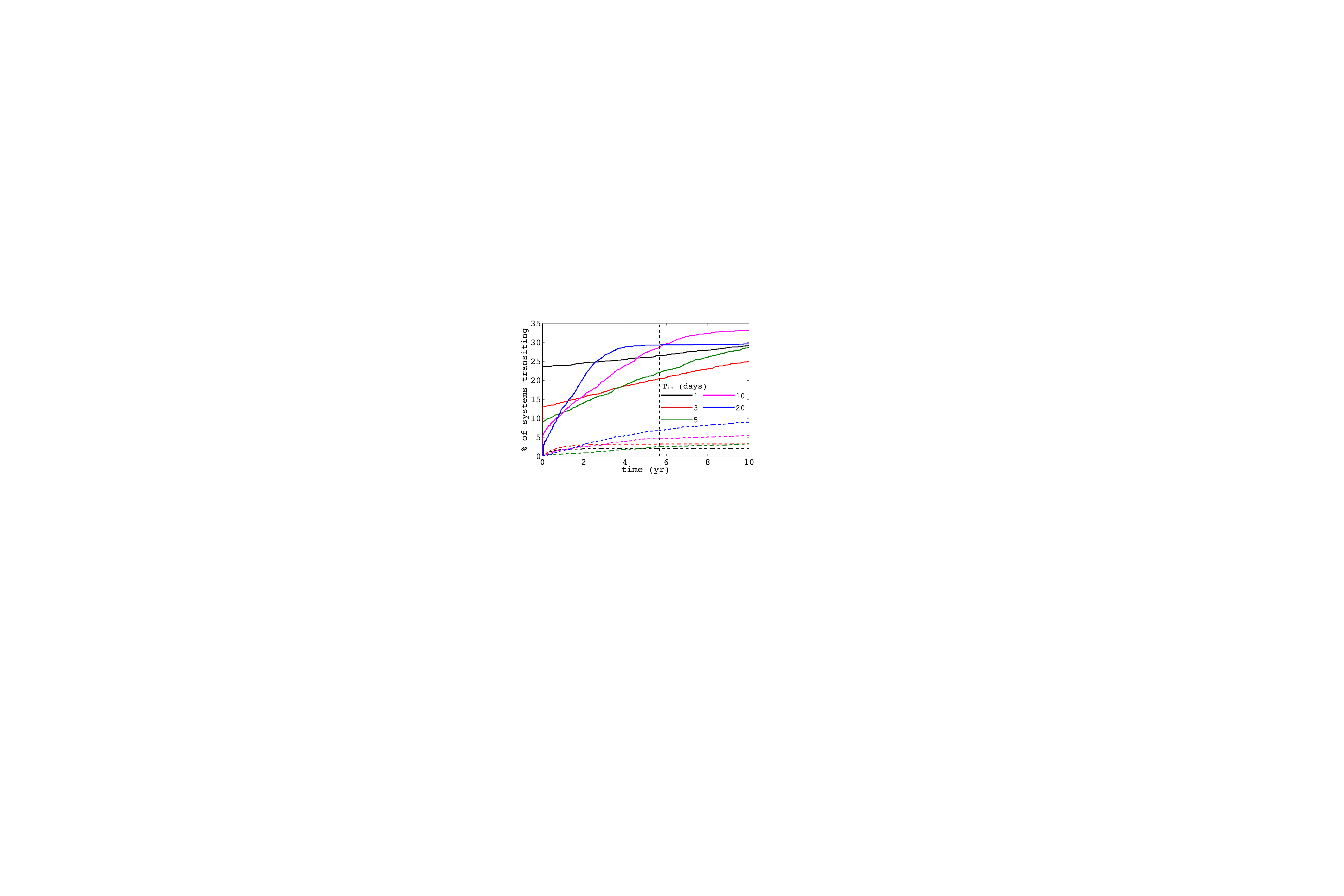}  
\caption{N-body simulations of transits on the host star (solid lines) and companion star (dashed lines) for a circumprimary planet in a binary with Solar and half-Solar stars. In all simulations $T_{\rm out}=100$ d (binary period) and $\Delta I = 15^{\circ}$. Different colours indicate different planet periods $T_{\rm in}$: 1 d (black), 3 d (red), 5 d (green), 10 d  (magenta) and 20 d (blue).}
\label{fig:inner_period_test}
\end{center}  
\end{figure} 

Whilst increasing $T_{\rm out}$ was shown to always decrease transit probabilities on both stars, for $T_{\rm in}$ it is less straight forward. Equation~\ref{eq:prob_host_body} predicts $P_{\rm host}$ to decrease as $T_{\rm in}$ increases, whilst conversely Eq.~\ref{eq:prob_comp_body} predicts a higher $P_{\rm comp}$ for longer $T_{\rm in}$. These predictions are true in the long term, but as we know, the rate of  precession is key in transits occurring within an observable timeframe, and $T_{\rm prec} \propto 1/T_{\rm in}$. N-body simulations are shown in Fig.~\ref{fig:inner_period_test} for a circumprimary planet with different $T_{\rm in}$ and a fixed $T_{\rm out}=100$ d. For transits on the host star we see there is a balance: short-period planets are geometrically closer to the star (that's good) but precess slower (that's bad). A short observing time ($< 1$ year) favours short $T_{\rm in}$ but after 10 years the 10 day planet transits the most. Only the 20 day planet completes a full precession period. For transits on the companion star it is simpler: longer period planets are more likely to transit, as they are not only geometrically favourable but precess faster.

\subsubsection{Effect of the mutual inclination}

\begin{figure}  
\begin{center}  
\large{Circumprimary planet simulations}
\includegraphics[width=0.49\textwidth]{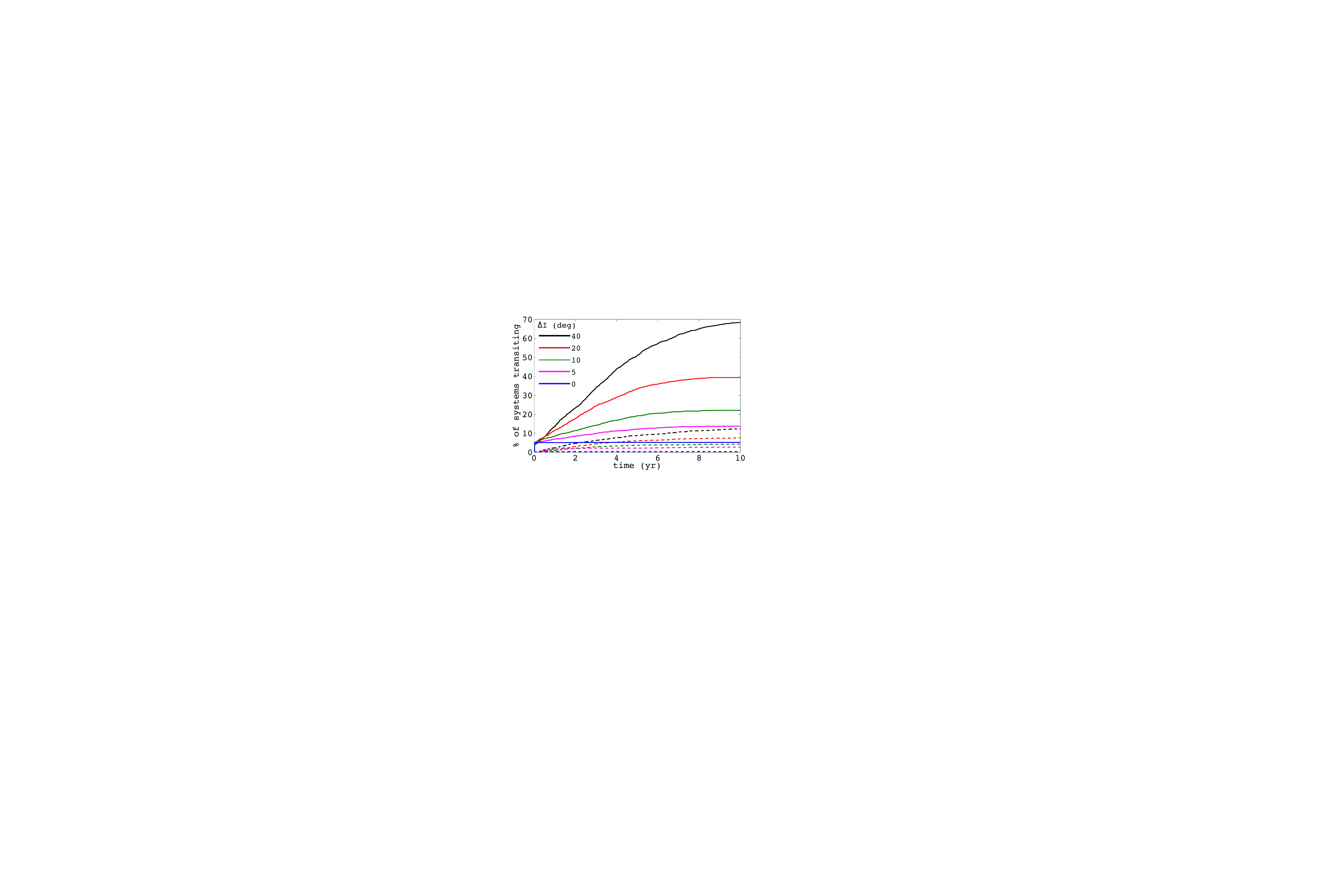}  
\caption{N-body simulations of transits on the host star (solid lines) and companion star (dashed lines) for a circumprimary planet in a binary with Solar and half-Solar stars. In all simulations $T_{\rm in}=11$ d (planet period) and $T_{\rm out}=100$ d (binary period). Different colours indicate different mutual inclinations $\Delta I$: 40$^{\circ}$ (black), 20$^{\circ}$ (red), 10$^{\circ}$ (green), 5$^{\circ}$  (magenta) and 0$^{\circ}$ (blue).}
\label{fig:example_mutualincl}
\end{center}  
\end{figure}

A key result of this work is the importance of $\Delta I$. In Fig.~\ref{fig:example_mutualincl} the relationship between the transit probability and $\Delta I$ is shown for an 11 day circumprimary planet in a 100 day binary with $\Delta I$ between 0 and 40$^{\circ}$. For transits on the host star the initial value is the same irrespectively of $\Delta I$, but the higher the $\Delta I$ the higher the transit probability is over time. Even a mere 5$^{\circ}$ mutual inclination can double the transit probability on the host star within just four years. For coplanar planets the transit probability is static and minimised due to the lack of precession. For transits on the companion star we see a similar trend of $\Delta I$ aiding transit probabilities.


\begin{figure}  
\begin{center}  
\large{Exomoon simulations}
\includegraphics[width=0.49\textwidth]{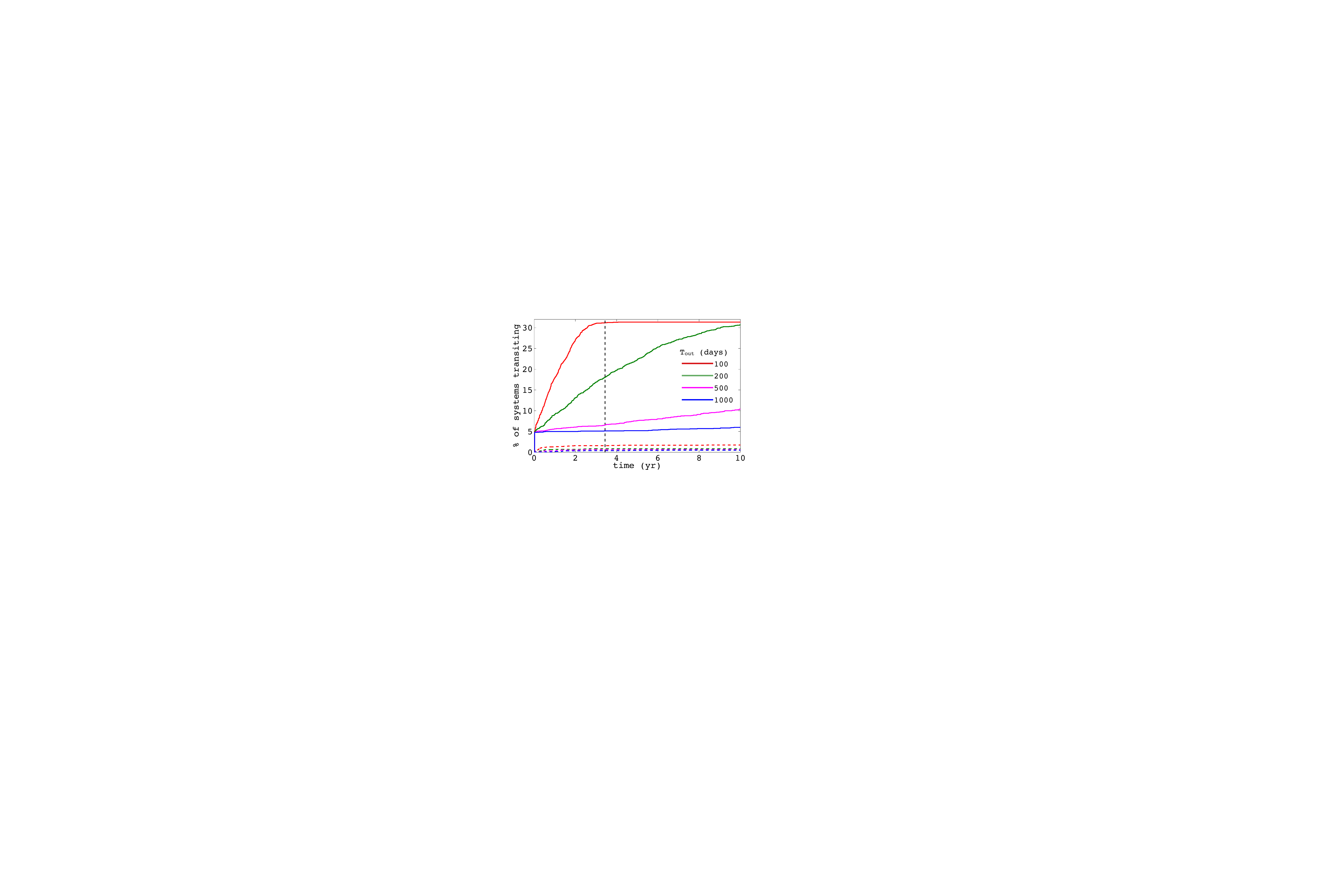}  
\caption{N-body simulations of transits on the host planet (solid lines) and companion star (dashed lines) for an exomoon around a Jupiter planet that is in turn orbiting a Solar star. In all simulations $T_{\rm in}=11$ d (moon period) and $\Delta I=15^{\circ}$. Different colours indicate different planet periods $T_{\rm out}$: 100 d (red), 200 d (green), 500 d  (magenta) and 1000 d (blue).}
\label{fig:example_exomoon}
\end{center}  
\end{figure}


\subsection{Exomoons}\label{sec:examples}

In Fig.~\ref{fig:example_exomoon} we run similar simulations to Fig.~\ref{fig:outer_period_test} but with a Jupiter host and Solar companion. Transits on the host body  (the planet) occur at a slightly faster rate than the circumprimary planet case, owing to a shorter $T_{\rm prec}$, but transits on the companion body (the star) are significantly reduced in the exomoon case. This is because for the same $T_{\rm in}=11$ d period $a_{\rm in}$ is much smaller for an exomoon (0.0095 AU) than for the circumsecondary planet (0.0968 AU).  This reduced probability on the star is unfortunate since those are the events likely to be actually observable.

\subsection{Observational signature}
\label{subsec:observational_signature}

\begin{figure}  
\begin{center}  
	\begin{subfigure}[b]{0.49\textwidth}
		\includegraphics[width=\textwidth]{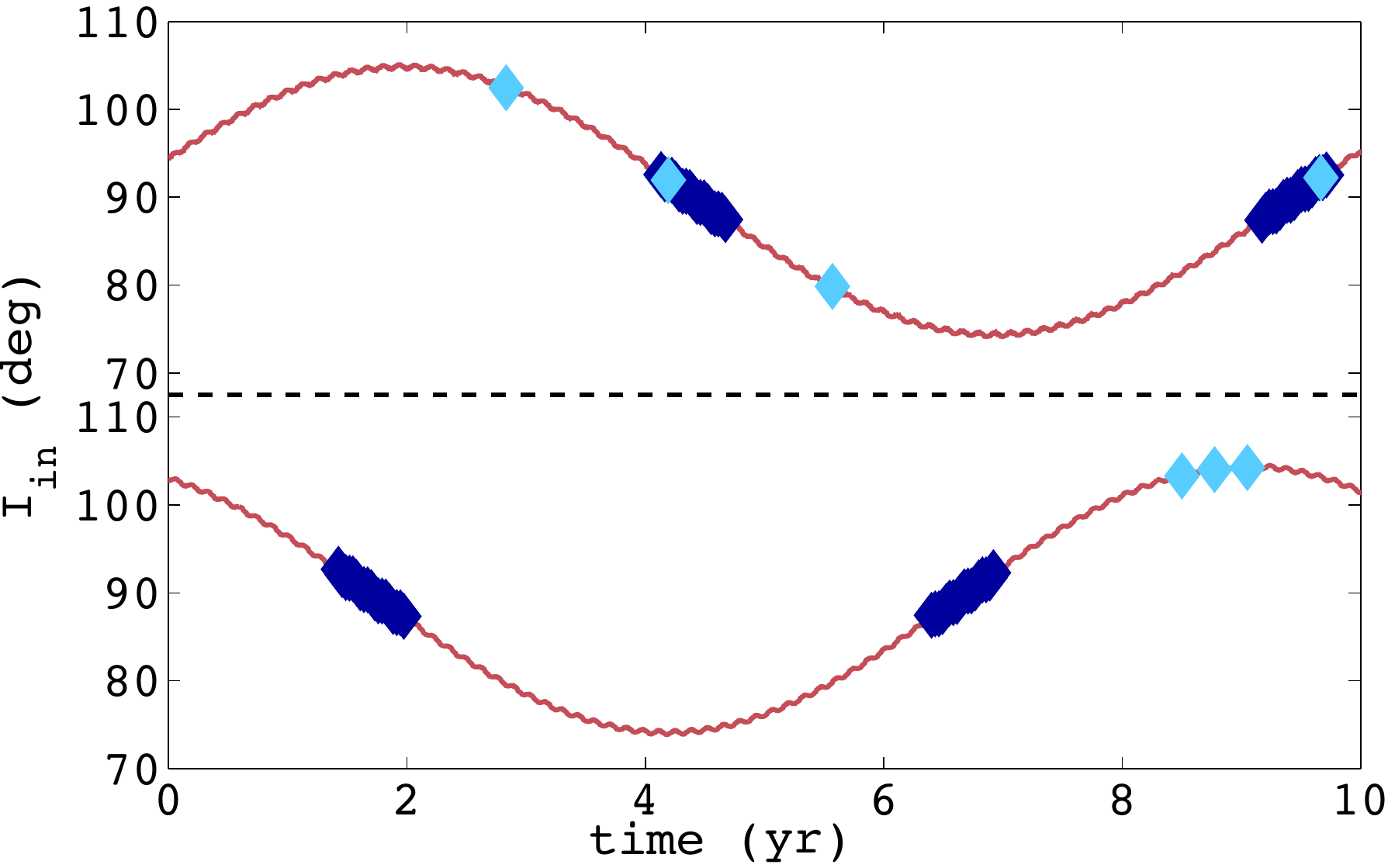}  
		\caption{Time variation of $I_{\rm in}$ for two example circumprimary planets from the simulations in Fig.~\ref{fig:outer_period_test} with $\Delta I = 15^{\circ}$, $T_{\rm in}=11$ d and $T_{\rm out}=100$ d, where the binary is known to eclipse. Transits on the host and companion star are denoted by dark blue and light blue diamonds, respectively. The dark blue diamonds look merged together because on the host there are concentrated batches of $\sim 18$ transits separated by only $T_{\rm in}$ whenever $I_{\rm in} \approx 90^{\circ}$.}
	\end{subfigure}
	\begin{subfigure}[b]{0.49\textwidth}
		\includegraphics[width=\textwidth]{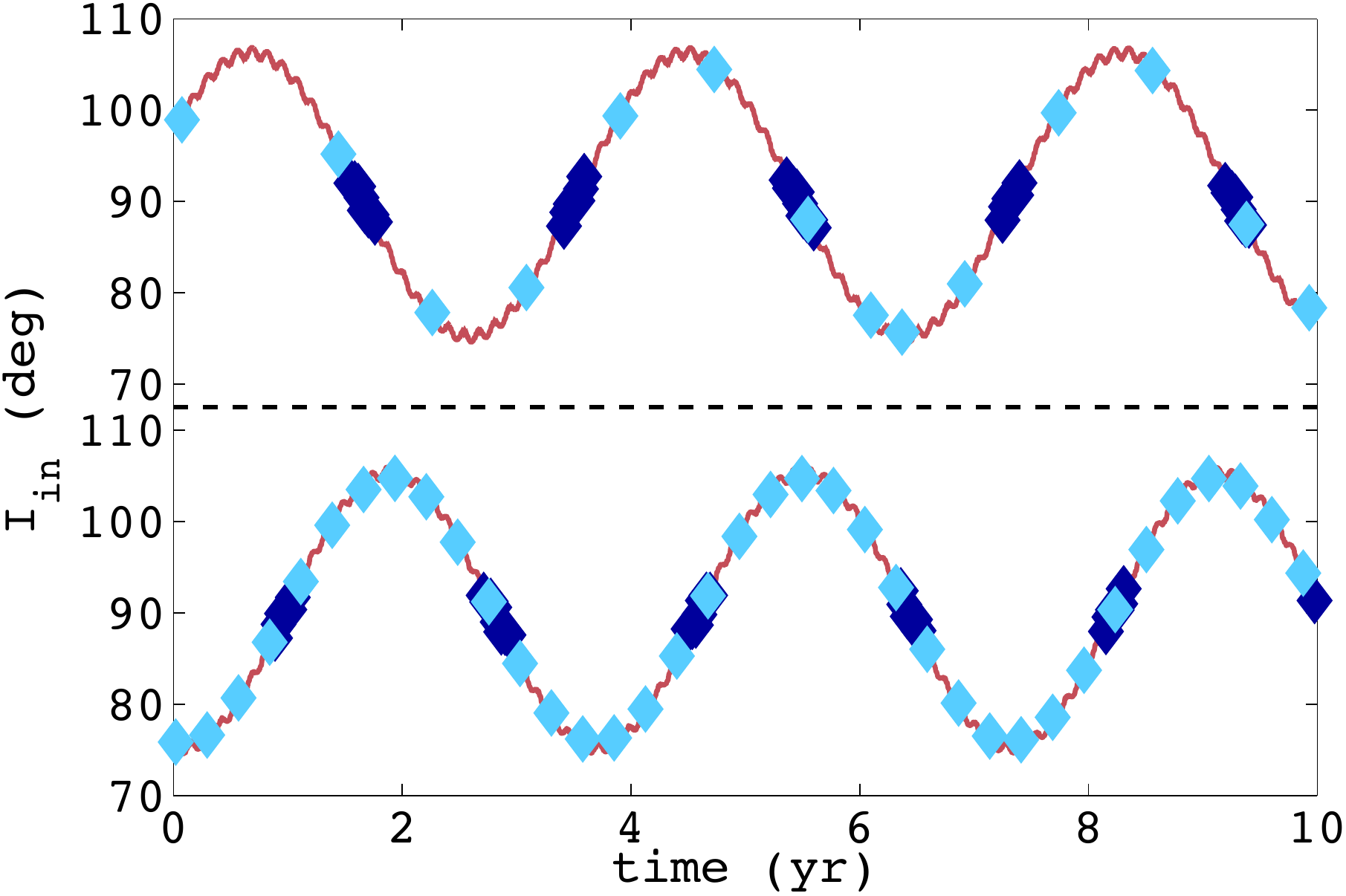}  	
		\caption{The same as (a) but for two exomoon simulations taken from Fig.~\ref{fig:example_exomoon}  with $\Delta I = 15^{\circ}$, $T_{\rm in}=11$ d and $T_{\rm out}=100$ d. In both examples the planet hosting the moon is known to transit the star. Like in (a) transits on the host (the planet) come in large consecutive bunches when $I_{\rm in}\approx90^{\circ}$ but transits on the companion (the star) are much more numerous than in (a), owing to a smaller $a_{\rm in}$ for the same $T_{\rm in}$.}
	\end{subfigure}
	\caption{}
\label{fig:observational_signature}  
\end{center}  
\end{figure}

Two examples from the  circumprimary simulations in Fig.~\ref{fig:outer_period_test} with $\Delta I = 15^{\circ}$, $T_{\rm in}=11$ d and $T_{\rm out}=100$ d are used to illustrate the timing of transits on the host and companion star, the results of which are shown in Fig.~\ref{fig:observational_signature}a. Both of these binaries eclipse. The variation of $I_{\rm in}$ is shown to follow Eq.~\ref{eq:I_in}, except for tiny additional short-period fluctuations. Consecutive sequences of $\sim 18$ transits on the host occur when $I_{\rm in}$ is near $90^{\circ}$.  This transient signal may be missed if search algorithms are  not tuned to analyse the light curve in segments.  Transits on the companion are rarer, sparsely distributed and can occur at at any $I_{\rm in}$. In both examples there are only three transits on the companion, which was typical of the 2000 simulations run; planets that transited the companion did so an average of 3.8 times. These will be harder to detect and characterise, in a similar way to inclined circumbinary planets \citep{martin15,kostov14,martin17}.

 We next show in Fig.~\ref{fig:observational_signature}b the comparable results for two of the exomoon simulations from Fig.~\ref{fig:example_exomoon}, where again we use $\Delta I = 15^{\circ}$, $T_{\rm in}=11$ d and$T_{\rm out}=100$ d. The amplitude of the variation of $I_{\rm in}$ is the same as for the circumprimary planets in Fig.~\ref{fig:observational_signature}a, but the precession period is slightly shorter as expected from Eq.~\ref{eq:precession_period}. Transits on the host planet again occur in consecutive blocks when the moon's inclination reaches $90^{\circ}$, but as has been stressed throughout this paper this is likely not an observable phenomenon. Transits on the companion star, however, occur at any $I_{\rm in}$ and in these two examples in Fig.~\ref{fig:observational_signature}b are very numerous: 14 in the top simulation and 37 in the bottom, the latter corresponding to one moon transit every orbit of its planet. These two examples in Fig.~\ref{fig:observational_signature}b are representative of the 2000 simulations run, as moons that were observed to transit the companion star on average did so 21 times, much more than the average of 3.8 in the circumprimary example. The difference is that for the exomoon $a_{\rm in}=0.0095$ is $\sim10$ times smaller than for the circumprimary planet where $a_{\rm in}=0.0968$ AU, simply due to Kepler's laws. We therefore have an interesting trade-off: widening the satellite's orbit {\it increases} increases its chance of transitability on the companion, and hence the long-term transit probability according to Eq.~\ref{eq:prob_comp_body}, but doing so {\it decreases} the efficiency of transitability, making the satellite harder to detect.

\begin{figure}  
\begin{center}  
	\begin{subfigure}[b]{0.49\textwidth}
		\includegraphics[width=\textwidth]{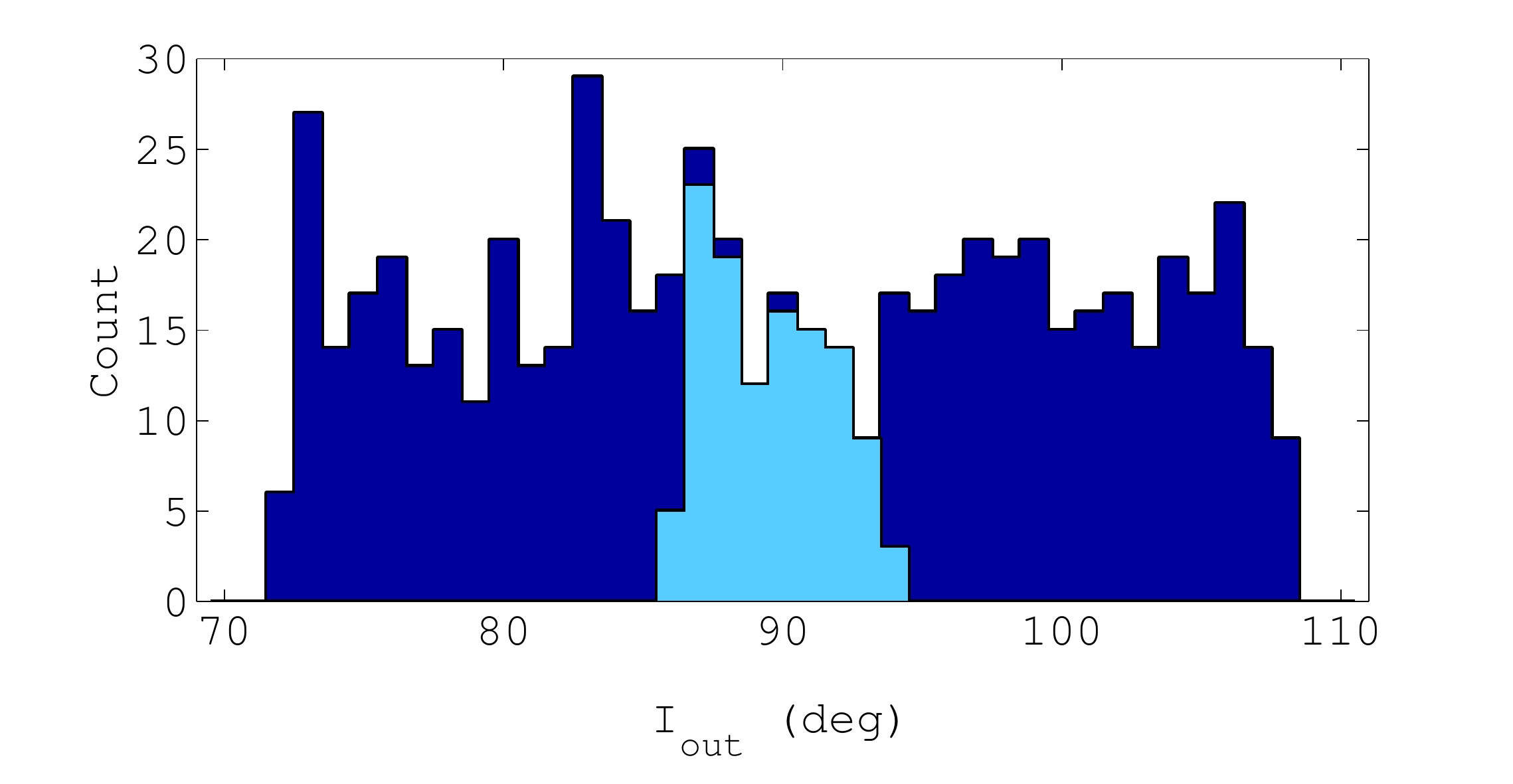}  
		\caption{Histogram of $I_{\rm out}$ for all transiting circumprimary planets from the simulations in Fig.~\ref{fig:outer_period_test} with $\Delta I = 15^{\circ}$, $T_{\rm in}=11$ d and $T_{\rm out}=100$ d. The dark blue histogram is for planets transiting the host star and the light blue histogram is for transits on the companion star. The histogram bin widths are 1$^{\circ}$. When $I_{\rm out}\approx90^{\circ}$ the planet is orbiting in an eclipsing binary.}
	\end{subfigure}
	\begin{subfigure}[b]{0.49\textwidth}
		\includegraphics[width=\textwidth]{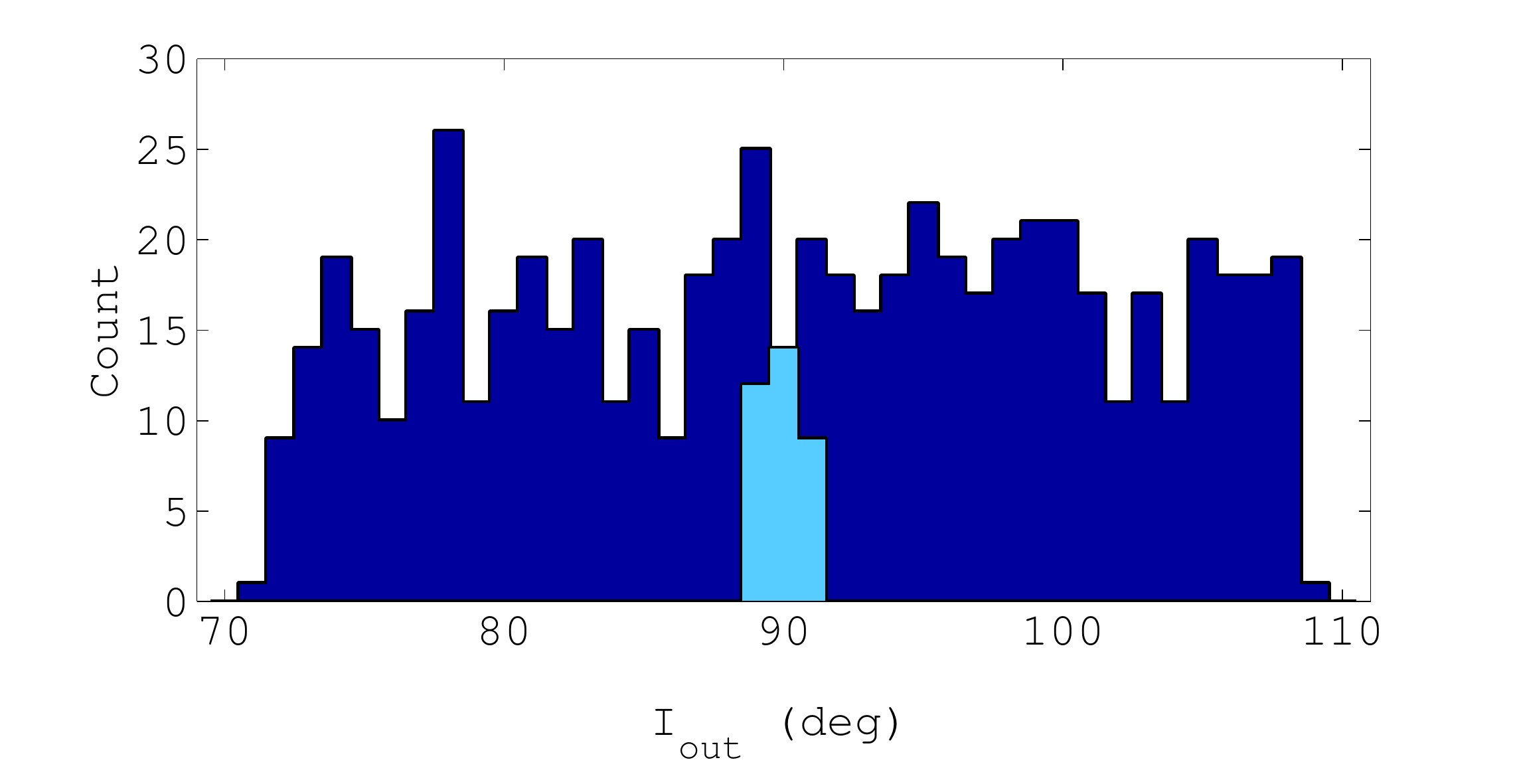}  	
		\caption{The same as (a) but for exomoon simulations from Fig.~\ref{fig:example_exomoon}  with $\Delta I = 15^{\circ}$, $T_{\rm in}=11$ d and $T_{\rm out}=100$ d. When $I_{\rm out}\approx90^{\circ}$ the exomoon is orbiting a transiting planet.}
	\end{subfigure}
	\caption{}
\label{fig:EB_test}  
\end{center}  
\end{figure} 

\subsection{Advantage of eclipsing binaries  and exomoons around transiting planets}

 The results throughout this paper are calculated assuming istroropically distributed systems (a uniform distribution of $\cos I_{\rm out}$). What happens if we instead know that $I_{\rm out}\approx 90^{\circ}$, i.e. a circumstellar planet within an eclipsing binary or an exomoon around a transiting planet? From Fig.~\ref{fig:geometry} it is evident that observers on the celestial sphere who see the companion body in front of the host body are almost guaranteed to be within both of the hatched regions designating transits on the host and companion,  and hence we have a favourable bias.

 To further demonstrate this, we take transiting circumprimary planets from the simulations in Fig.~\ref{fig:outer_period_test}, with $\Delta I=15^{\circ}$, $T_{\rm out}=100$ d and  $T_{\rm in} = 11$ d, and plot a histogram of $I_{\rm out}$, shown in Fig.~\ref{fig:EB_test}a for both host (dark blue) and companion transits (light blue). For transits on both bodies there is seen to be a range of $I_{\rm out}$ for which transits are permissible, centred around 90$^{\circ}$. This range is much larger for transits on the host than the companion. Within this range the histogram appears roughly flat, but more thorough simulations would be needed to prove this. 

In the 2000 simulations there were 618 planets transiting the host star and 95\% of them did so on non-eclipsing binaries. This is due to the wide range of $I_{\rm out}$ permitting transits and the simple fact that most binaries do not eclipse. This number is reduced to 67\% for transits on the companion. Whilst eclipsing binaries therefore are not a necessity for transits, they are highly beneficial as in this small sample transits occurred on the host and companion at a rate of 100 and 97\%, respectively, within the simulated 10 yr.

Even though eclipsing binaries are biased towards short periods, and one may not expect circumstellar planets in say a $<100$ d binary, the  {\it Kepler} mission discovered 167 eclipsing binaries with periods of 100 d and longer\footnote{Original sample in \citet{prsa11} and up to date catalog available at \url{http://keplerebs.villanova.edu/}.}. {\it If} they do host planets there is a very high chance of them transiting,  albeit with a complicated signature.

 In Fig.~\ref{fig:EB_test}b we produce the same histogram but for exomoon simulations from Fig.~\ref{fig:example_exomoon}, again with $\Delta I=15^{\circ}$, $T_{\rm out}=100$ d and  $T_{\rm in} = 11$ d. Compared with the circumprimary case the range of $I_{\rm out}$ permitting transits on the host is roughly the same. However, transits on the companion star only occur in a much narrower range of $I_{\rm out}$ around $90^{\circ}$. Consequently, only 38\% of exomoons transited the star did so when the planet {\it did not} transit. This suggests that for searches for transiting exomoons it is recommended but not required that the planet transits.


 Finally, another advantage of a satellite that orbits an eclipsing binary or a transiting planet is that those eclipses/transits would also be identified by a continuous transit survey like {\it Kepler}, {\it TESS} or {\it PLATO} and can provide complementary information  like eclipse/transit timing variations \citep{sartoretti99,kipping09,kipping15,oshagh16}.

%
%

\section{Conclusion}
\label{sec:conclusion}

 The equations and N-body simulations in this paper have demonstrated that for  circumstellar planets in close binaries and exomoons in planetary systems,  orbital precession enhances transit probabilities and makes them inherently time-dependent. The transit probability of the satellite (exomoon or circumstellar planet) on the host body (for the exomoon that is its planet) rises roughly linearly over a precession period before plateauing at the value calculated in this paper. This probability may be as high as tens of per cent, largely as a function of the mutual inclination, and may be reached over a precession period as short as 5-10 yr for outer orbits of 100-200 d. Transits come and go, exhibiting consecutive sequences whenever the satellite's inclination is near $90^{\circ}$. Transits on the companion body (for the exomoon that is its star) are systematically less likely but still have a time-dependence, which rises even slightly beyond the precession period. They may occur at any point during the satellite's precession period but depending on the orbital parameters may be in a sparse sequence and difficult to characterise.

This odd transit signature draws parallels with inclined circumbinary planets \citep{martin15,martin17}, such as the  marginally inclined ($4^{\circ}$) Kepler-413 \citep{kostov14}. Even if large circumstellar planets in close binaries have individually significant transits, they may be missed by automated detection pipelines that rely on transit regularity. For exomoons there is the added difficulty of the transits themselves generally being undetectably small, and a clever dynamical phase-folding would be necessary to recover a signal.

Searches for circumstellar planets in close binaries are enhanced (and sometimes guaranteed) if the binary is known to eclipse, particularly if transits are desired on both stars, but this is not a necessity. The sample of almost 3,000 eclipsing binaries found throughout the {\it Kepler} mission, with more to come particularly from {\it PLATO}, provides hope of finding such planets {\it if} they exist. To find exomoons transit their star it is highly recommended that the planet transits too. 

Finally, we note that this work may be applied to eclipses/transits of close triple star systems and similar-mass binary planets.

\section{Acknowledgements}

Thank you to my PhD supervisor Stephane Udry for constant support and discussions, Amaury Triaud for our circumbinary work that inspired this paper, Rosemary Mardling for  the orbital precession timescale, and Jean Schneider for the \citet{flammarion74} reference.  This work was aided by the thoughtful comments of an anonymous referee. Finally, I thank the financial support of the Swiss National Science Foundation over the past four years.

\end{document}